\documentclass[12pt,preprint]{aastex63}
\usepackage{color}
\usepackage[normalem]{ulem}  % \sout{old text} for strikeout

\begin{document}

\title{Properties of binary components and remnant in GW170817 using 
equations of state in finite temperature field theory models}

\correspondingauthor{Debades Bandyopadhyay}
\email{debades.bandyopadhyay@saha.ac.in}

\author[0000-0001-6946-9227]{Shriya Soma}
\affil{Frankfurt Institute for Advanced Studies (FIAS), Ruth-Moufang-Strasse 1,
D60438 Frankfurt am Main, Germany}
\affil{Astroparticle Physics and Cosmology Division, Saha Institute of Nuclear 
Physics (SINP), 1/AF Bidhannagar, Kolkata-700064, India}
\email{soma@fias.uni-frankfurt.de}
\author[0000-0003-0616-4367]{Debades Bandyopadhyay} 
\affil{Frankfurt Institute for Advanced Studies (FIAS), Ruth-Moufang-Strasse 1,
D60438 Frankfurt am Main, Germany}
\affil{Astroparticle Physics and Cosmology Division, Saha Institute of Nuclear 
Physics (SINP), 1/AF Bidhannagar, Kolkata-700064, India}
\affil{Homi Bhabha National Institute (HBNI), Training School Complex, 
Anushaktinagar, Mumbai-400094, India}

\begin{abstract}
We investigate gross properties of binary components and remnant 
in GW170817 using equations of state within the finite 
temperature field theoretical models. We also adopt finite temperature 
equations of state in the density dependent hadron field theory for this study. 
Properties of binary components are studied 
using zero temperature equations of state. Particularly, we investigate tidal 
deformabilities and radii of binary components in
light of GW170817. An analytical expression relating the radii and the combined
tidal deformability is obtained for binary neutron star masses in the range 
$1.1M_{\odot}\lesssim M\lesssim 1.6 M_{\odot}$. The upper bound on the tidal 
deformability gives the upper bound on the neutron star radius as 
13 km. Next, the role of finite temperature on the remnant in GW170817 is
explored. In this case, we investigate the gravitational and baryon mass, 
radius, Kepler frequency and moment of inertia of the rigidly rotating remnant 
for different equations of state at fixed entropy per baryon. The remnant 
radius is enlarged due to thermal effects compared with the zero temperature 
case. Consequently, it is found that the Kepler frequency is much lower at 
higher entropy per baryon than that of the case at zero temperature. 
These findings are consistent with the results found in the literature. 

\end{abstract}


\section{Introduction}
The discovery of the first pulsar in 1967 by Jocelyn Bell \citep{bell} not only 
led to the 
quest for the densest form of matter in compact astrophysical objects in this 
observable universe, but also prepared the stage to probe the 
dense matter through gravitational waves when two neutron stars in the binary 
collided 50 years later on 17 August, 2017. This heralded in the gravitational 
wave astronomy. The phase evolution of gravitational wave signal in late stage 
inspirals encodes the information about the equation of state (EoS) of dense 
matter through tidal deformation in one neutron star due to the other in a 
binary neutron star merger. This tidal deformability parameter was extracted 
for the first time analysing the gravitational wave signal from the first 
binary neutron star merger GW170817 and a constraint was imposed on the EoS of 
dense matter \citep{ligo1,ligo2,ligo3,ligo4}. 

Neutron stars are unique laboratories for dense matter under extreme 
astrophysical conditions.  
After the pulsar discovery, the study of dense matter in terrestrial 
laboratories gained momentum. Though Quantum Chromodynamics predicts a 
very rich phase structure of dense matter, only a small region of it can be 
probed in laboratories. It has been possible to produce hot and dense matter 
in heavy ion collisions. Future experimental facilities such as the 
International Facility for Antiproton Ion Beam research (FAIR) at GSI, Germany
would create hot ( a few tens of MeV) and dense matter (a few times the 
normal nuclear matter density) relevant for newly born neutron stars. The study
of hot and dense matter in laboratories enriches our knowledge about the 
modification of hadron properties in dense medium, the properties of strange 
matter involving hyperons, (anti)kaons and a phase transition from hadrons to
quarks. This empirical information from heavy ion collisions is highly 
valuable in understanding dense matter in neutron star interior and binary
neutron star mergers \citep{magic}. This shows that there is a strong interplay
between the dense matter produced in heavy ion collisions and formed in 
neutron star interior.     

It is a well known fact that measured masses, radii and moments of inertia are
the direct probes of dense matter in neutron star interior. Precise mass 
measurement of pulsars has been possible through the estimation of post 
Keplerian parameters such as Shapiro delay, using pulsar timing. This led to 
the discoveries of very massive neutron stars \citep{demo,anto}. 
A 2 M$_{\odot}$ neutron star was discovered for the first time employing this 
technique \citep{anto}. The most massive neutron star is 
the millisecond pulsar PSR0740+6620 with a mass 2.14$^{+0.10}_{-0.09}$ at 
68.3$\%$ credibility interval which has been reported this year by the North 
American Nanohertz Observatory for Gravitational Waves (NANOGrav) 
\citep{croma}. This sets the lower bound on a neutron star's maximum mass. On 
the other hand, it has been argued that an upper bound on the maximum mass of 
neutron stars could be
placed from electromagnetic observations if the remnant formed in neutron star 
merger event GW170817 had collapsed to a black hole in a second or so 
\citep{luci,metz,shapiro,shibata}. The predicted upper bound on the neutron 
star maximum mass is $\simeq 2.17$ M$_{\odot}$. However a higher upper bound 
$\sim 2.3$ M$_{\odot}$ has been reported recently \citep{enping}. Both bounds 
on the maximum mass of neutron stars put stringent conditions on the 
composition and EoS of dense matter in neutron star 
interiors. It has been found that the presence of non-nucleonic forms of matter
such as hyperons, kaon condensates and quarks makes the EoS softer, which might
be incompatible with the massive neutron star in most cases. 

The radius measurement of neutron stars is a difficult task. The Neutron Star
Composition Explorer (NICER) mission in the space station is devoted to the
estimation of neutron star radius by observing x-rays from rotation powered 
pulsars. The NICER has just published the results on the mass and radius 
of millisecond pulsar PSR J0030+0451 as 1.44$^{+0.15}_{-0.14}$ M$_{\odot}$ and 
13.02$^{+1.24}_{-1.06}$ km \citep{miller}. 
Riley et al. also reported very similar results of mass (1.34$^{+0.15}_{-0.16}$
M$_{\odot}$) and radius (12.71$^{+1.14}_{-1.19}$ km) in a different analysis
of the NICER results \citep{riley}.  On the other hand, the extracted value 
of the combined tidal 
deformability $70\leq \widetilde{\Lambda} \leq 720$ from GW170817 provides 
important information about radii of neutron stars involved in the binary 
neutron star merger event GW170817 \citep{ligo3,sch18,fattoyev,ozel,de,zhao} 
for the tidal deformability strongly correlates with the radius of a neutron 
star. All 
those calculations indicate that the radius of 1.4 M$_{\odot}$ neutron stars 
could be $9 \leq R/km \leq 14$. 

Valuable information about the EoS of neutron star matter might be imprinted
in the post merger gravitational wave signal from the remnant. The 
frequency of gravitational waves in the post merger phase could be as high as
a few kilo hertz. LIGO and VIRGO detectors are less sensitive to such high
frequencies. Consequently, no post merger signal was detected from the remnant
of GW170817 \citep{ligo5}. Electromagnetic observations provided whatever 
little information we have about the remnant in GW170817. It was argued 
that the 
amount of blue kilonova ejecta as observed in the electromagnetic counterpart of
GW170817 might not be compatible with a prompt collapse to a black hole. The 
fate of the remnant could be one of three possibilities - (i) the delayed
collapse of a hypermassive neutron star (HMNS) supported by differential 
rotation to a black hole in a second or so, (ii) a supramassive neutron star 
(SMNS) supported by rigid body rotation survives longer before 
collapsing to a black hole and (iii) a forever stable neutron star. All three 
possibilities depend on whether the mass of the remnant is above or below a
threshold mass as well as the maximum mass of the corresponding non-rotating 
neutron star. It is possible to predict certain properties of the remnant if 
it becomes a uniformly rotating object. Recently, the LIGO-VIRGO 
collaboration (LVC) predicted the moment of inertia and maximum rotation rate 
of the uniformly rotating remnant at the mass-shedding limit considering a 
large set of zero temperature equations of state (EoSs) \citep{ligo6}.   

The binary neutron star merger produces a remnant that is highly hot and 
dense \citep{seki,radice1}. This problem has been studied in numerical 
relativity simulations by different groups 
\citep{ott,shibata,radice2,radice3,giacomazzo,most}. These simulations predict
that the maximum temperature of the remnant could be $\sim 70$ MeV or more and 
the maximum density $\sim 5$ times the normal nuclear matter density. 
Such high baryon densities in the remnant might lead to the appearance of new 
degrees of freedom such as hyperons and quarks \citep{ott,most}. It also 
follows 
from the simulations that the entropy per baryon is $s_B \lesssim 2 k_B$ 
at the core of the remnant just after the merger whereas the bulk of the 
remnant outside the unshocked core has the entropy per baryon which is a few 
times $k_B$. Although there is a large spread in temperature and entropy 
initially, the conditions become homogeneous at a later time.
The early evolution of the remnant is driven by gravitational 
wave radiation over 10-20 ms followed by the later evolution due to 
viscosity on a time scale of $\sim 100$ ms and neutrino cooling in 2-3 s 
\citep{kiuchi,kenta,fuji,seki,radice2}. The magnetic field plays an 
important role on the evolution of the remnant. An effective viscosity is 
expected to be generated in the remnant through the magnetorotational 
instability (MRI). The other effect competing with the MRI is the magnetic 
winding during differential rotation of the remnant \citep{kenta}. Both effects
are responsible for transporting angular momentum and removing the differential 
rotation making it a rigidly rotating remnant \citep{giacomazzo}.  
It would be interesting to investigate
the thermal effects on the structures and Keplerian frequencies of a uniformly 
rotating remnant. The thermal effects on the remnant in BNS 
merger were earlier studied using EoSs at fixed temperatures \citep{kaplan}. In 
this work, our motivation is to investigate the properties of the remnant 
in the numerical library LORENE using EoSs at fixed entropy per baryon 
\citep{eric}.    

The paper is organised as follows. In Section 2, different field theory models 
of EoSs at zero and finite temperatures, as well as the numerical library  
LORENE for the study of rigidly rotating remnant are described. The results of 
our calculation are discussed in Section 3. Section 4 contains the summary and 
conclusions. 

\section{Methodology}
We adopt $beta$-equilibrated and charge neutral equations of state at zero and
finite temperatures constructed within the framework of relativistic mean 
field (RMF) models with and without density dependent (DD) couplings to compute
tidal deformability, structures of uniformly rotating neutron stars, moment of
inertia and Keplerian frequencies. Baryon-baryon interaction, in RMF models, is 
mediated by the exchange of $\sigma$, $\omega$ and $\rho$ mesons. Moreover,
hyperon-hyperon interaction is mediated by the exchange of $\phi$ mesons.    
Furthermore, we consider two classes of EoSs - unified and non-unified. In 
case of unified EoSs, the same nucleon-nucleon interaction of RMF models is 
employed in low and high density matter. For non-unified case, we use the RMF
model including non-linear $\sigma$ meson self-interaction terms to describe
the neutron star matter EoS in the core which is matched with the outer and 
inner crust EoS given by Baym-Pethick-Sutherland and Negele and Vautherin 
\citep{bps,neg}. 

On the other other hand, for unified EoS, an extended version of the nuclear 
statistical equilibrium takes care of the matter made of light and heavy 
nuclei, and unbound nucleons at low temperatures and below the saturation 
density \citep{hs1}. The interaction among unbound nucleons is dictated by RMF
models which are also used to describe the matter at high densities. We exploit
different parametrizations of RMF models such as DD2, SFHo, SFHx, TM1, TMA
\citep{typ10,rmp} for nuclear matter EoS, Banik, Hempel, Bandyopadhyay (BHB) 
$\Lambda$ hyperon EoS known as BHB$\Lambda \phi$ EoS \citep{bhb} and 
hadron-quark (Hybrid) EoS undergoing a first order phase transition from 
hadronic 
to quark matter \citep{weber}. We discuss the density dependent relativistic 
hadron (DDRH) field theory model at finite temperature along with other unified
EoSs in the following paragraphs in details.

\subsection{EoS in DDRH field theory model at finite temperature}

The starting point in this model is the Lagrangian density as given by 
\citep{typ10,bhb},  
\begin{eqnarray}
{\cal L}_B &=& \sum_{B=n,p,\Lambda} \bar\Psi_{B}\left(i\gamma_\mu{\partial^\mu}
- m_B
+ g_{\sigma B} \sigma - g_{\omega B} \gamma_\mu \omega^\mu 
- g_{\rho B} 
\gamma_\mu{\mbox{\boldmath $\tau$}}_B \cdot 
{\mbox{\boldmath $\rho$}}^\mu 
- g_{\phi B} \gamma_\mu \phi^\mu 
\right)\Psi_B\nonumber\\
&& + \frac{1}{2}\left( \partial_\mu \sigma\partial^\mu \sigma
- m_\sigma^2 \sigma^2\right) 
 -\frac{1}{4} \omega_{\mu\nu}\omega^{\mu\nu}
+\frac{1}{2}m_\omega^2 \omega_\mu \omega^\mu
- \frac{1}{4}{\mbox {\boldmath $\rho$}}_{\mu\nu} \cdot
{\mbox {\boldmath $\rho$}}^{\mu\nu}
+ \frac{1}{2}m_\rho^2 {\mbox {\boldmath $\rho$}}_\mu \cdot
{\mbox {\boldmath $\rho$}}^\mu \nonumber\\
&& -\frac{1}{4} \phi_{\mu\nu}\phi^{\mu\nu}
+\frac{1}{2}m_\phi^2 \phi_\mu \phi^\mu~.
\label{lagm}
\end{eqnarray}
Here $m_B$ is the bare baryon mass,  ${\mbox{\boldmath $\tau_{B}$}}$ is the 
isospin operator and $\Psi_B$ denotes the isospin multiplets for baryons. The
parametrization of this model involving only nucleons is known as DD2 
\citep{typ10}. We extend the Lagrangian to include $\Lambda$ hyperons and 
hyperon-hyperon interaction is mediated by $\phi$ mesons. This model is denoted
as BHB$\Lambda \phi$ \citep{bhb}.

The grand canonical partition function in the mean field approximation 
can be written as,
\begin{eqnarray}
ln Z_{B} &=& {\beta}V[-\frac{1}{2}m_\sigma^2 \sigma^2
+ \frac{1}{2} m_\omega^2 \omega_0^2 
+ \frac{1}{2} m_\rho^2 \rho_{03}^2  
+ \frac{1}{2} m_\phi^2 \phi_0^2 
+ \Sigma^r \sum_{B=n,p,\Lambda} n_B]
\nonumber \\
&& + 2V \sum_{i=n,p,\Lambda} \int \frac{d^3 k}{(2\pi)^3} 
[ln(1 + e^{-\beta(E^* - \nu_i)}) +
ln(1 + e^{-\beta(E^* + \nu_i)})] ~,  
\end{eqnarray}
where the temperature is defined as $\beta = 1/T$,
$E^* = \sqrt{(k^2 + m_B^{*2})}$ and effective baryon mass 
$m_B^* = m_B - g_{\sigma} {\sigma}$. The chemical potential of i-th baryon is
given by 
\begin{equation}
\mu_i = \nu_i + g_{\omega B} \omega_0 + g_{\rho B} \tau_{3B} \rho_{03} 
+ g_{\phi B} \phi_0 + \Sigma^r~,
\label{vec}
\end{equation}
and the rearrangement term which takes care of many-body correlations, has the 
form,
\begin{equation}
\Sigma^r=\sum_{B=n,p,\Lambda}[-\frac {\partial g_{\sigma B}} {\partial n_B} 
\sigma n_B^s 
+ \frac { \partial g_{\omega B}}{\partial n_B} \omega_0 n_B
+ \frac { \partial g_{\rho B}}{\partial n_B}\tau_{3B} 
\rho_{03} n_B
+ \frac { \partial g_{\phi B}}{\partial n_B} 
\phi_0 n_B]~.
\label{rea}
\end{equation}
The total grand canonical partition function of the system is 
$Z = Z_{B} Z_{L}$ where 
$Z_{L}$ denotes the grand canonical partition function for 
non-interacting leptons.

We obtain the equations of motion for meson fields by extremising 
$Z_{B}$. Furthermore, we can compute all thermodynamic quantities 
of baryonic matter using $Z_{B}$.
The baryon pressure is
written as $P = T V^{-1} lnZ_{B}$ and the energy density of baryons is,
\begin{eqnarray}
\epsilon &=& \frac{1}{2}m_\sigma^2 \sigma^2
+ \frac{1}{2} m_\omega^2 \omega_0^2 
+ \frac{1}{2} m_\rho^2 \rho_{03}^2  
+ \frac{1}{2} m_\phi^2 \phi_0^2 
\nonumber \\
&& + 2 \sum_{i=n,p,\Lambda} \int \frac{d^3 k}{(2\pi)^3} E^* 
\left({\frac{1}{e^{\beta(E^*-\nu_i)} 
+ 1}} + {\frac{1}{e^{\beta(E^*+\nu_i)} + 1}}\right)~.  
\end{eqnarray}
The number density of i$(=n,p,\Lambda$)-th baryon is
$n_i = 2 \int \frac{d^3k}{(2\pi)^3} \left({\frac{1}{e^{\beta(E^*-\nu_i)} 
+ 1}} - {\frac{1}{e^{\beta(E^*+\nu_i)} + 1}}\right)$.
The scalar density for baryon $B$ ($n_B^s$) is
\begin{eqnarray}
n_B^s &=& 
2 \int \frac{d^3 k}{(2\pi)^3} \frac{m_B^*}{E^*} 
\left({\frac{1}{e^{\beta(E^*-\nu_B)} 
+ 1}} + {\frac{1}{e^{\beta(E^*+\nu_B)} + 1}}\right) ~.
\end{eqnarray}
The entropy density of baryons follows from the relation
$S = \beta \left(\epsilon + P - \sum_{i=n,p,\Lambda} \mu_i n_i \right)$.
and the entropy density per baryon is $s = {S}/{n_b}$ where $n_b$ is
the total baryon density.

Nucleon-meson couplings in the DDRH model are density dependent. The DD2 
parameter set of nucleon-meson couplings is used to describe the nuclear matter
properties \citep{wol,typ10}. The functional forms of density dependent
couplings $g_{\sigma N}$ and $g{\omega N}$ are given by,
\begin{eqnarray}
g_{\alpha N} = g_{\alpha N} (n_0) f_{\alpha} (x)~,\nonumber\\ 
f_{\alpha} (n_b/n_0) = a_{\alpha} \frac{1+b_{\alpha}(x+d_{\alpha})^2}
{1+c_{\alpha} (x +d_{\alpha})^2}~,
\label{coef}
\end{eqnarray}
where $n_0$ is the saturation density, $\alpha = \sigma, \omega$ and
$x = n_b/n_0$. For $\rho$ mesons, we have,
\begin{eqnarray}
g_{\rho N} = g_{\rho N} (n_0) exp{[-a_{\rho} (x - 1)]}~. 
\label{rhoc}
\end{eqnarray}
Coefficients in both equations, saturation density, nucleon-meson couplings at 
the saturation density, mass of $\sigma$ mesons are obtained by fitting the 
properties of finite nuclei \citep{typ10}. The properties of symmetric nuclear 
matter at the saturation density ($n_0=0.149065$ fm$^{-3}$) are consistent with
the experimental values \citep{rmp}. The symmetry energy (32.73 MeV) and its 
density slope (57.94 MeV) are in consonance with experimental findings and 
observations of neutron stars \citep{jim,tews1,lona}. Furthermore, DD2 EoS is 
reasonably compatible with that of pure neutron matter obtained in the chiral 
effective field theory \citep{rmp,heb}. 

On the other hand, $\Lambda$ hyperon-vector meson 
couplings are determined from the SU(6) symmetry relations \citep{dov,sch} and
$\Lambda$ hyperon - scalar meson coupling is extracted from the hypernuclei 
data. We consider $\Lambda$ hyperon potential depth -30 MeV in normal nuclear 
matter \citep{mil,mar,sch92}. Hyperon-meson couplings are taken from 
Ref.\citep{bhb}. Both DD2 and BHB$\Lambda \phi$ EoSs are publicly available 
from CompOSE and being widely used for supernovae and neutron star merger 
simulations \citep{composemanual}.  

\subsection{EoS in non-liner Walecka model at finite temperature}
Here we introduce unified Steiner, Fischer and Hempel (SFH) EoSs based on 
NSE model for matter below the saturation density and non-linear Walecka model 
with additional meson couplings \citep{hs1,stein05}. The non-linear Walecka 
model with cross meson terms is given by \citep{stein05}, 
$$ \mathcal{L} = \sum_{B=n,p} \bar{\Psi}_{B} (i\gamma _{\mu} \partial ^{\mu} - m_B + g_{\sigma B} \sigma - g_{\omega B} \gamma_{\mu} \omega ^{\mu}-\frac{1}{2}g_{\rho B} \gamma _{\mu}{\mbox{\boldmath $\tau$}}_B \cdot {\mbox{\boldmath $\rho$}}^{\mu})\Psi_B $$
$$ + \frac{1}{2} (\partial _{\mu} \sigma \partial^{\mu}\sigma - m_{\sigma}^2 \sigma ^2) 
- \frac{1}{4} \omega_{\mu\nu} \omega^{\mu\nu} + \frac{1}{2} m_\omega ^2 \omega_{\mu} \omega^{\mu} 
- \frac{1}{4}{\mbox{\boldmath $\rho$}}_{\mu\nu} \cdot {\mbox{\boldmath $\rho$}}^{\mu\nu} + \frac{1}{2} m_\rho ^2 {\mbox{\boldmath $\rho$}}_{\mu} \cdot {\mbox{\boldmath $\rho$}}^{\mu} - U(\sigma)$$
\begin{equation}
+ {\frac{\kappa}{24}} g_{\omega B}^4 (\omega^{\mu}\omega_{\mu})^2 + 
{\frac{\lambda}{24}} g_{\rho B}^4 ({\mbox{\boldmath $\rho$}}^{\mu} \cdot 
{\mbox{\boldmath $\rho$}}_{\mu})^2 
+ g_{\rho B}^2 f(\sigma,\omega^{\mu}\omega_{\mu}) 
{\mbox{\boldmath $\rho$}}^{\mu} \cdot {\mbox{\boldmath $\rho$}}_{\mu}~.   
\end{equation}

${\tau}_B$ is the isospin operator, and $U(\sigma)$ represents the self interaction terms, and can be expanded as
\begin{eqnarray}
U(\sigma) = {\frac{\zeta}{6}}(g_{\sigma B}\sigma)^3+
{\frac{\xi}{24}}(g_{\sigma B}\sigma)^4~,
\end{eqnarray}
and
\begin{eqnarray}
f(\sigma,{\omega^{\mu} \omega_{\mu}}) = \sum_{1}^{6} a_i \sigma^i + \sum_{1}^{3}
b_j (\omega^{\mu} \omega_{\mu})^j~.
\end{eqnarray}
There are 17 parameters in this model. These provide enough freedom to fine 
tune the low and high density part of the isospin sector independently 
\citep{stein13}. These two EoSs are known as SFHo and SFHx where `o` stands for
optimal and `x` stands for extremal. In SFHo, the most probable mass-radius
curve of Ref. \citep{stein10} was fitted whereas the radius of low mass neutron
stars was minimised in SFHx model resulting in low value (23.18 MeV) for the 
density slope of the symmetry energy at the saturation density \citep{stein13}.   
If we neglect last two terms of the Lagrangian density given by Eq.(9), it 
reduces to the same Lagrangian density of TM1 and TMA EoS models with different
parameter sets \citep{toki,toki95}. In 
this case too, a unified EoS was constructed based on the NSE model for the low
density matter and the Lagrangian density without last two terms in Eq. (9) for 
nucleon-nucleon interaction for low as well as high density matter 
\citep{hem12}. 

\subsection{Hybrid EoS at zero temperature}
We also consider an EoS undergoing a first order phase transition from hadronic
to quark matter governed by the Gibbs phase rules. In this case, the hadronic
matter is described by the DD2 Lagrangian density of Eq. (\ref{lagm}) extended
to include all hyperons of $1/2$-spin baryon octet and $\Delta$ resonance 
\citep{weber}. The three flavour quark matter is described by the nonlocal 
extension of the Nambu-Jona-Lasino model as introduced in Ref.{\citep{weber}}. 
This hybrid EoS is calculated at temperature T=0. This hybrid EoS will be 
exploited to compute properties of binary components, but not 
the remnant.  

\subsection{Rapidly Rotating Remnant}
Once the differential rotation of the remnant is removed over the 
effective viscous and magnetic winding time scale $\sim 100$ ms, the 
remnant becomes a rigidly rotating body. A stationary, axisymmetric spacetime 
is assumed for the study of this rigidly rotating remnant formed in GW170817. 

Stationary and axisymmetric rapidly rotating star models within 
general relativity was numerically studied in $3+1$ dimensional space plus time 
framework \citep{BGSM}. Here the 4-dimensional spacetime manifold is foliated 
into a family of non-intersecting space-like hypersurfaces $\Sigma_t$ 
parametrized by coordinate time $t$. Defining three spatial coordinates 
($x^i$) on each hypersurface, one can write the line element in terms of lapse 
function $N$ and shift vector ($\beta^i$),
\begin{equation}
ds^2 = -N^2 dt^2 + {\gamma}_{ij} (dx^i + \beta^i dt) (dx^j + \beta^j dt)~,
\end{equation}    
where $\gamma_{ij}$ is the 3-metric on each $\Sigma_t$.

Coordinates are chosen for this problem from the consideration of spacetime
symmetries and foliation in 3+1 framework. It is assumed here that the 
spacetime is stationary, axisymmetric and asymptotically flat. This points
to the fact that there are two commuting Killing vector fields 
(${\bf {e_0}} = \frac{\partial}{\partial t}$ and 
${\bf{e_3}} = \frac{\partial}{\partial \phi}$) in the chosen 
coordinates ($x^0 = t$, $x^1$, $x^2$, $x^3=\phi$). Remaining two coordinates
($x^1=r, x^2=\theta$) are chosen as spherical.  Furthermore, 
$\beta^r$ = $\beta^{\theta}$ =0, $\gamma_{r\phi} = \gamma_{\theta\phi}=0$.
Working in a quasi-isotropic gauge which makes $\gamma_{r\theta} = 0$,  
the line element reduces to the form \citep{marq}, 
\begin{equation}
ds^2 = - N^2 dt^2 + A^2(dr^2+r^2 d\theta^2) + B^2 r^2 \sin^2{\theta} (d\phi^2 - {\beta}^{\phi} dt)^2~,
\label{eq:metric}
\end{equation}
where metric potentials $N,\beta^{\phi},A,B$ depend on coordinates $r$ 
and 
$\theta$. Finally four gravitational field equations were obtained as a set of 
four coupled elliptic partial differential equations involving 
energy-momentum tensor in source terms \citep{BGSM}.

The matter is described by the energy-momentum tensor of a perfect fluid,
\begin{eqnarray}
T^{\mu \nu} &=& (\varepsilon + P) u^{\mu} u^{\nu} + P g^{\mu \nu}~.
\label{emtensor}
\end{eqnarray}

The fluid log-enthalpy is
\begin{equation}
 H = ln \left(\frac{\varepsilon + P}{n m_B}\right)~,
\end{equation}
where $n$ and $m_B$ are baryon density and rest mass, respectively.

The equation of the fluid equilibrium follows from the conservation of 
energy-momentum tensor
\begin{equation}
H(r,\theta) + ln N - \ln \Gamma (r,\theta) = \frac{T e^{-H}}{m_B} \partial_i s
- u_{\phi}u^t \partial_i \Omega, 
\label{equi} 
\end{equation}
where $\Gamma$ is the Lorentz factor of the fluid with respect to the Eulerian
observer, $s$ is the entropy per baryon in Boltzmann unit. As we consider only
rigid rotation i.e. $\Omega=$ constant, the last term vanishes. It is shown 
that the equilibrium equation (\ref{equi}) finally reduces to the zero 
temperature expression \citep{marq}   
\begin{equation}
H(r,\theta) + ln N - \ln \Gamma (r,\theta) = const~. 
\label{eq:first-integral} 
\end{equation}

We use the numerical library LORENE which implements the above formulation 
\citep{eric}.

\begin{deluxetable}{rccccccccc}[ht!]
\tablecolumns{17}
\tablewidth{0pc}
\tablecaption{
The saturation properties of nuclear matter
such as saturation density ($n_0$), dimensionless effective nucleon mass, 
binding energy (BE), incompressibility (K), symmetry energy (S), and density 
slope of symmetry energy (L) are obtained using the different parameter 
sets. Maximum mass of non-rotating neutron stars and the corresponding baryon
mass are also mentioned here. Experimental values of nuclear matter 
properties at the saturation density quoted in the last row are taken from 
Ref.\citep{rmp,stone,horn,perego}.} 
\tablehead{
EoS& $n_0$&$m^{*}/m$&BE&K&S&L&M$_{max}$&M$_B$\\
 & ($fm^{-3}$) & &(MeV)&(MeV)& (MeV) &(MeV)&(M$_{\odot}$)&(M$_{\odot}$)}
\startdata
DD2 &0.1491&0.56&16.02&243.0& 31.67&55.04&2.42&2.89\\  
BHB$\Lambda \phi$ &0.1491&0.56&16.02&243.0& 31.67&55.04&2.1&2.43\\  
SFHo &0.1583&0.76&16.19&245.4&31.57&47.10&2.06&2.43\\  
SFHx &0.1602&0.72&16.16&238.8&28.67&23.18&2.13&2.53\\  
TM1&0.1455&0.63&16.31&281.6&36.95&110.99&2.21&2.30\\  
TMA&0.1472&0.64&16.03&318.2&30.66& 90.14&2.02&2.30\\  
G230a&0.153&0.78&16.30&230.0&32.50&89.76&2.01&2.31\\  
G230b&0.153&0.70&16.30&230.0&32.50&94.46&2.33&2.75\\  
G240a&0.153&0.78&16.30&240.0&32.50&89.70&2.02&2.75\\  
G240b&0.153&0.70&16.30&240.0&32.50&94.39&2.34&2.75\\  
G300a&0.153&0.78&16.30&300.0&32.50&89.33&2.08&2.40\\  
G300b&0.153&0.70&16.30&300.0&32.50&93.94&2.36&2.78\\  
Hybrid &0.1491&0.56&16.02&243.0& 31.67&55.04&2.05&2.39\\  
Exp. &0.15-0.16&0.55-0.75&16.00&220-315&29.00-31.70&45.00-61.90&-&-\\
\hline
\enddata
\label{table}
\end{deluxetable}

\begin{figure}[ht!]
\epsscale{0.60}
\plotone{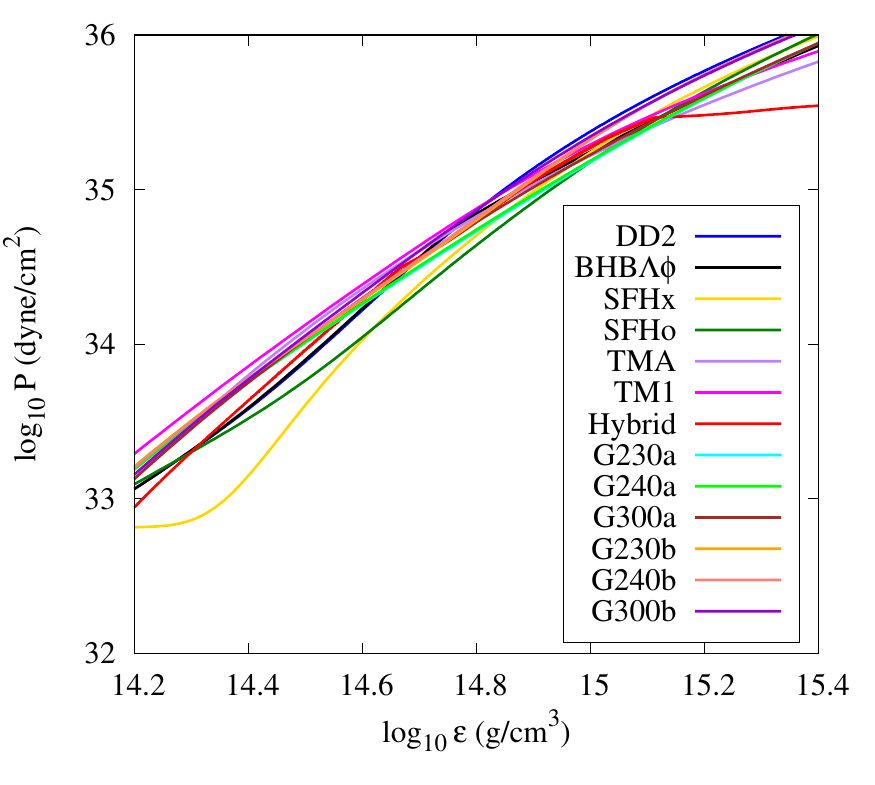}
\caption{Pressure as a function of energy density is plotted for various EoS 
models at zero temperature.}
\end{figure}
\begin{figure}[hb!]
\epsscale{0.60}
\plotone{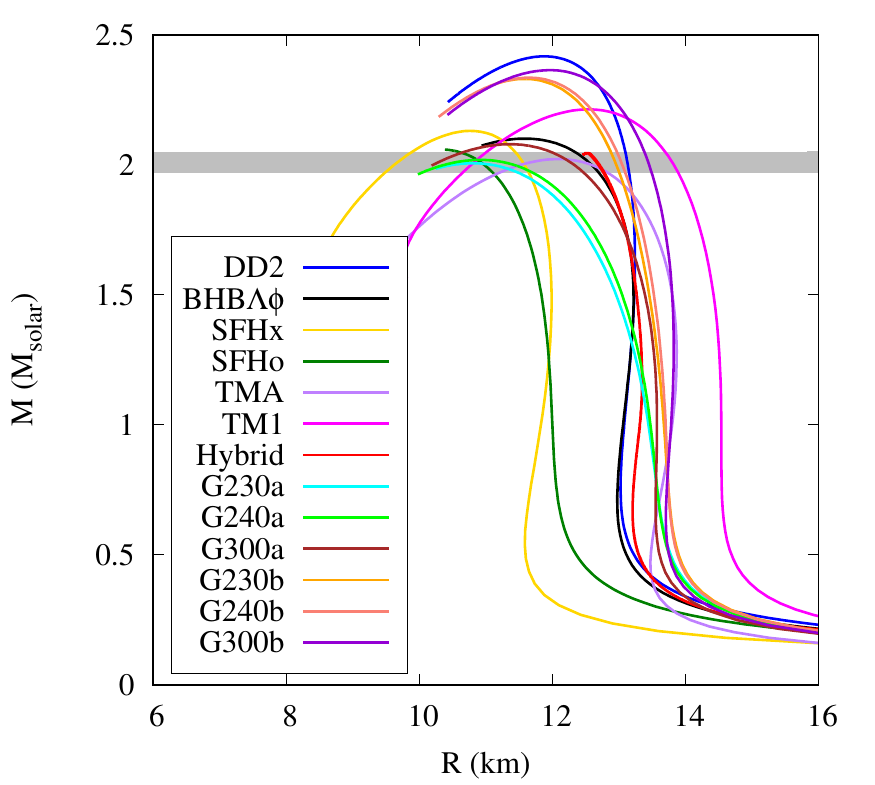}
\caption{Mass-Radius relationships are plotted for different EoSs. The grey
horizontal band denotes the bound on the neutron star maximum mass.}
\end{figure}
\section{Results and Discussion}       
We open our discussion with zero temperature EoSs and their role on 
tidal deformabilities and radii of binary components in GW170817. 
The saturation properties
of various EoSs used here are recorded in Table 1 along the maximum 
gravitational masses and corresponding baryonic masses. We denote the Walecka 
model with self-interaction terms of only $\sigma$ mesons given by Eq. (9) as
Glendenning (G) followed by incompressibility and a, b for two values of
effective masses, i.e. 0.78 and 0.7 respectively. Empirical values of 
nuclear
matter properties are reported in the last row of Table 1. The range of values
of incompressibility of nuclear matter at the saturation density is reported in
Ref. \citep{rmp,stone}. It was demonstrated that the value of nucleon effective
mass $0.55 \leq m^*/m \leq 0.75$ led to the physical solution for pure 
neutron matter which was compatible with the chiral effective field theory
\citep{horn}. Tews et al. reported
new bounds on the symmetry energy and its slope \citep{tews1}. It is evident 
from the table that the symmetry energy and its slope of several EoS models 
such as SFHx, TMA and Glendenning are in tension with experimental values, new 
bounds of Tews et al. and the state of the art calculations in the chiral 
effective field theory \citep{rmp,tews1,lona}. The low value of L in SFHx EoS 
model was obtained in an attempt to minimise the radii of low mass neutron 
stars \citep{stein13}. Figure 1 shows EoSs
constructed within the framework of different models as described in section 2.
Two kinks in the hybrid 
EoS represent the start and end of the mixed phase. The hybrid EoS becomes 
softer after the mixed phase ends. The low density part of the SFHx EoS 
deviates significantly from other EoSs. Furthermore,
it is noted that the pressure for SFHx around the saturation density remains 
constant. This kind of behaviour was also 
noted in the SFHx EoS of pure neutron matter \citep{fis}. We also plot
results of Glendenning EoS models. 
The mass-radius relationships of these EoSs are exhibited in Fig. 2. 
The grey band indicates the accurately measured pulsar mass of 
2.01$\pm 0.04$ M$_{\odot}$. 
We observe that
the lower effective mass leads to higher maximum 
mass of neutron stars than that of higher values of incompressibility for
Glendenning EoS models.   
Among all EoSs, the DD2 EoS results in the 
highest maximum mass neutron star. 
\begin{figure}
\epsscale{0.60}
\plotone{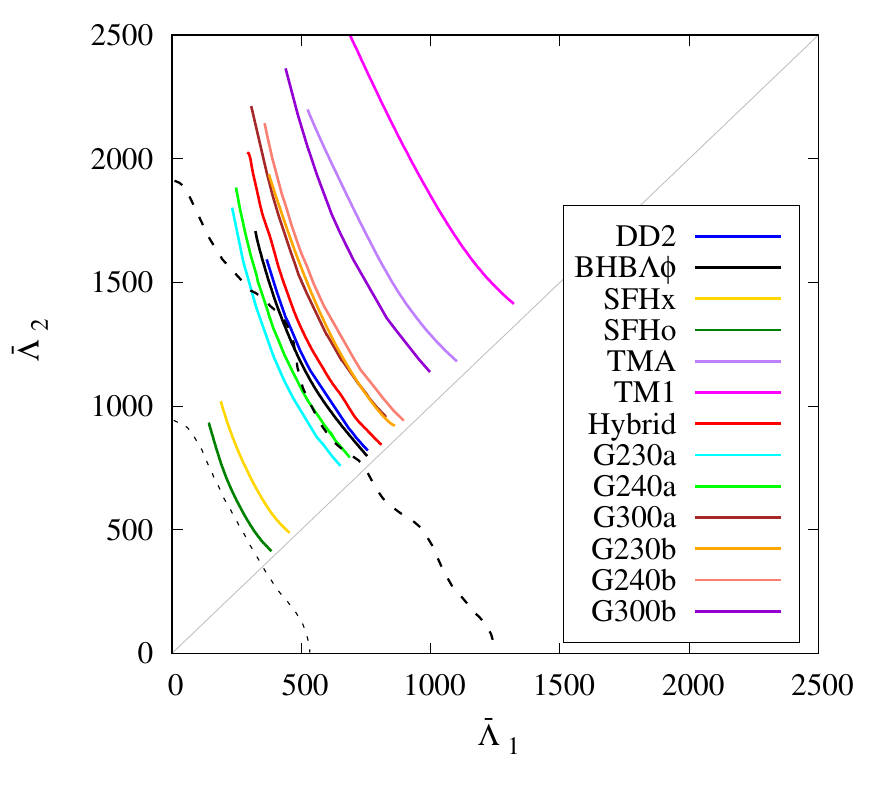}
\caption{${\bar{\Lambda}}_1$ is plotted with ${\bar{\Lambda}}_2$. Different EoSs
are overlaid on this plot. The dotted and dashed lines denote 50$\%$ and 
90$\%$ confidence levels extracted from gravitational wave signal of GW170817. }
\end{figure}
Figure 3 shows the plot of dimensionless tidal deformabilities 
${\bar{\Lambda}}_1$ and ${\bar{\Lambda}}_2$ corresponding to binary
components with masses $m_1$ and $m_2$, respectively. We apply the same zero 
temperature EoSs for the calculations of tidal deformabilities. The dotted and
dashed lines denote 50$\%$ and 90$\%$ confidence levels as obtained from the 
analysis of gravitational wave signal of GW170817 \citep{ligo1}. As one 
approaches from the top right corner to the bottom left corner, the compactness
increases. Equations of State predicting less compact neutron stars and
lying outside 90$\%$ confidence level are ruled out by GW170817,  whereas EoSs 
such as SFHo, SFHx, G230a and G240a are allowed and BHB$\Lambda \phi$ is 
marginally allowed. 
As neutron star masses involved in GW170817
ranged from 1.17 - 1.6 M$_{\odot}$, the combined tidal deformability probes only
a narrow density regime in neutron stars, whereas the maximum mass of a 
non-rotating neutron star compatible with the 2 M$_{\odot}$ pulsar is sensitive
to the EoS over a broader range of density from the density of crust to several
times the saturation density in the core.    

We know that the tidal deformability is closely related to the radius of a 
neutron star as evident from the expression,
\begin{eqnarray}
\bar{\Lambda} = \frac{2}{3}k_2 \Big{(} \frac{R}{M}\Big{)}^5~.
\end{eqnarray}
Different groups have already exploited the knowledge of tidal deformability in
GW170817 to estimate radii of neutron stars. An analytical approach was
prescribed to estimate the radius of a 1.4 M$_{\odot}$ neutron star relating 
the value of tidal deformability obtained from GW17017 \citep{zhao}. We extend 
this prescription to estimate the radius of a neutron star in the mass range   
$1.1M_{\odot}\lesssim M\lesssim 1.6 M_{\odot}$. 

\begin{figure}[ht!]
\epsscale{0.60}
\plotone{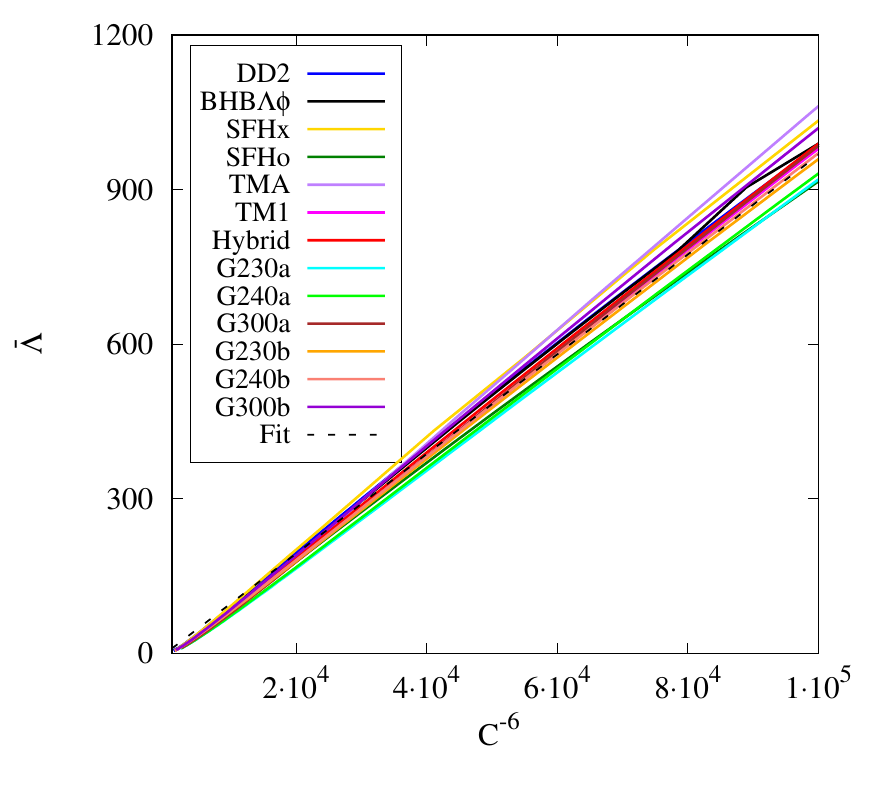}
\caption{Dimensionless tidal deformability is plotted with $C^{-6}$ where 
$C(=M/R)$ is the compactness.}
\end{figure}
It is noted that $\bar{\Lambda} \varpropto C ^{-6}$ because 
$k_2 \varpropto C$, where $C = {M}/{R}$, for a large collection of 
EoSs. We plot this relation in Fig. 4 for different EoSs. A fit to the curves  
with $\bar{\Lambda}= a C^{-6}$ yields the value of $a \simeq 0.00967$. 

The combined tidal deformability is defined as 
\begin{eqnarray}
\widetilde{\Lambda}  = \frac{16}{13} \frac{[(m_1 +12 m_2)m_1^4 \bar{\Lambda} _1 + (m_2 +12 m_1)m_2^4 \bar{\Lambda} _2] }{(m_1 + m_2)^5}~,
\end{eqnarray}
where $\bar{\Lambda} _1$ and $\bar{\Lambda} _2$ are the dimensionless tidal 
deformabilities of binary neutron star masses $m_1$ and $m_2$, respectively. 
Using the $\bar{\Lambda}= a C^{-6}$ relation and assuming radii of neutron
stars in the mass range $1.1M_{\odot}\lesssim M\lesssim 1.6 M_{\odot}$
nearly equal i.e. $R_1 \simeq R_2 \simeq R$, we get 
\begin{eqnarray}
{\widetilde{\Lambda}} = \frac{16a}{13} \times\frac{1}{(1+q)^5}  \Big{\{} (1+12q)\Big{(}\frac{R}{m_1} \Big{)} ^6 + q^4(q+12)\Big{(}\frac{R}{m_2}\Big{)} ^6\Big{\}}
\end{eqnarray}
$$ =\frac{16a}{13} \times\frac{1}{(1+q)^5} \times \Big{(}\frac{R}{m_1}\Big{)}^6 
\Big{\{} \frac{(1+12q)}{m_1^6} + \frac{ (q+12)}{q^2 m_1^6}\Big{\}}.$$

Writing $m_1$ in terms of chirp mass by using, 
${\cal{M}} = m_1q^{3/5}/(1+q)^{1/5}$,
$${\widetilde{\Lambda}} = \frac{16a}{13} \times \Big{(}\frac{R}{\cal{M}}\Big{)}^6\times\frac{q^{18/5}}{(1+q)^{31/5}} \times 
\big{[} q^2 + 12q^3 + q+12\big{]}.$$
Finally we get,
\begin{eqnarray}
{\widetilde{\Lambda}} = \frac{16a}{13} \times \Big{(}\frac{R}{\cal{M}}\Big{)}^6\times\frac{q^{8/5}}{(1+q)^{26/5}} \times \big{[} 12- 11q +12q^2 \big{]}~. 
\end{eqnarray}
It is noted that there is a weak dependence on 
mass ratio in the combined tidal deformability \citep{ozel,radice4,sajad}.
We investigate 
the extrema of ${\widetilde{\Lambda}}$ by taking the derivative with respect 
to $q$ at a fixed chirp mass and obtain
\begin{eqnarray}
\Big{(}\frac{\partial {\widetilde{\Lambda}}}{\partial q} \Big{)}_{\cal{M}} = {\widetilde{\Lambda}} \times \frac{1-q}{5q(1+q)} \times \Big{[}\frac{96-263q+96q^2}{12- 11q +12q^2}\Big{]}~.
\end{eqnarray}

It immediately follows that roots of this derivative are $q=1$ and $q=0.43346$.
We consider $m_2^{min} \sim 1M_{\odot}$ and $m_1^{max}\sim2M_{\odot}$ and this
implies that the relevant range of $q$ is clearly $\geq 0.5$, whereas for 
GW170817, $q \geq 0.7$ \citep{ligo1}. It may be concluded from Eq. (21) that 
the dependence of the combined tidal deformability on chirp mass
$\cal{M}$ is similar to that of $\bar{\Lambda}$ on $M$, i.e;
\begin{eqnarray}
{\widetilde{\Lambda}} = a'\Big{(}\frac{R}{\cal{M}}\Big{)}^6~.
\end{eqnarray}

In order to find $a'$, we can make a substitution of $q =0.7$ and $q=1$ in Eq. 
(21). This gives $a'\simeq 0.0042$ for $q=1$ and 
$a' \simeq 0.0043$ for $q=0.7$. As expected, the value of $a'$ changes 
negligibly for the entire relevant range of $q$ for GW170817.
\begin{figure}[ht!]
\epsscale{0.60}
\plotone{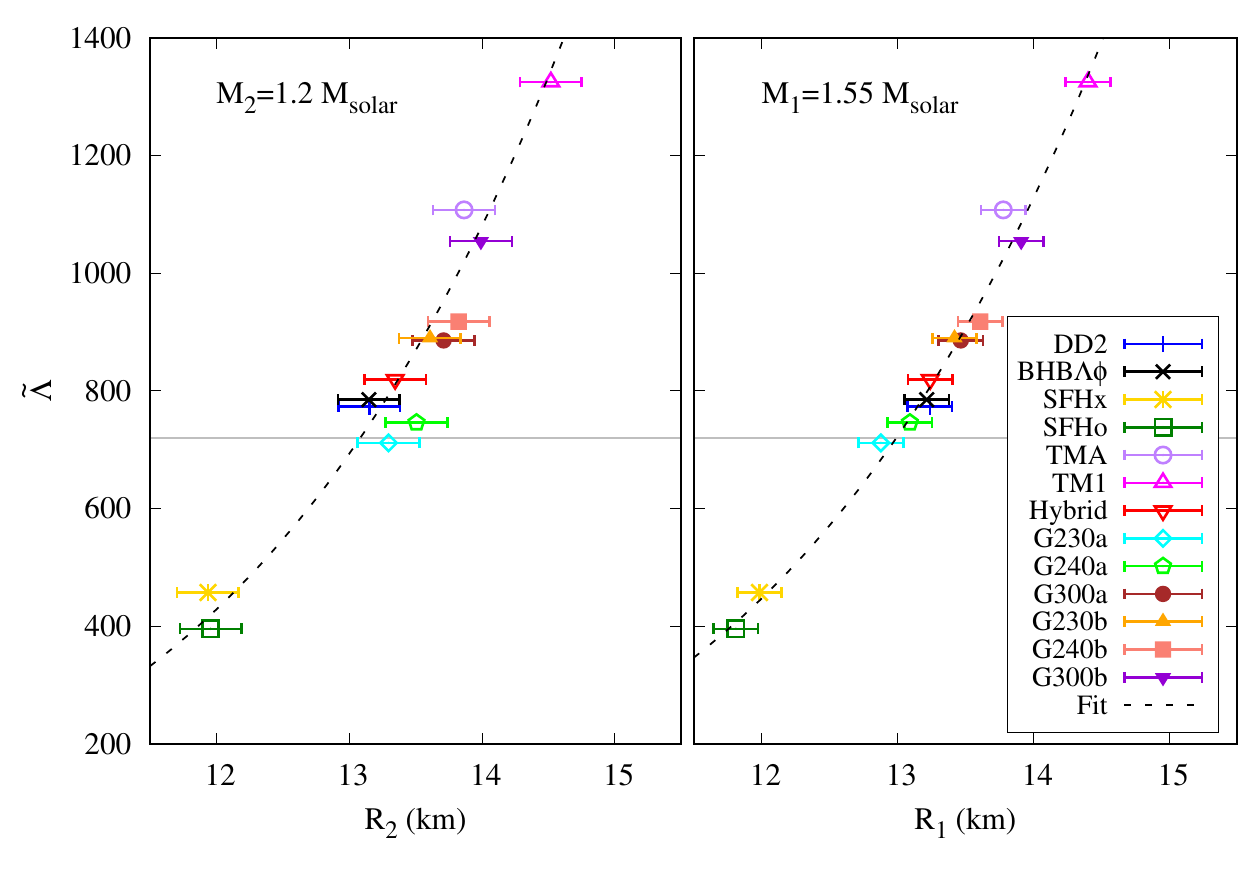}
\caption{The combined tidal deformability $\tilde{\Lambda}$ is plotted with
radius $R_1$ of primary component mass 1.55 M$_{\odot}$ in the right panel and
radius $R_2$ of secondary component mass 1.2 M$_{\odot}$ in the left panel. 
Symbols with colours represent different EoSs. The horizontal line in both 
panels is the upper bound on the combined tidal deformability.}
\end{figure}

We can now find the radii of the binary masses which lie in the range 
$1.17 - 1.6 M_{\odot}$ for GW170817 \citep{ligo3}.
\begin{eqnarray}
R = 3.669\times \frac{\cal{M} }{M_{\odot}}\times {\widetilde{\Lambda}}^{1/6}~.
\end{eqnarray}
For the chirp mass ${\cal{M}} = 1.188$ for GW170817, it becomes
\begin{eqnarray}
R = 4.36 \times {\widetilde{\Lambda}}^{1/6}~. 
\label{rad}
\end{eqnarray}
If we plug-in the upper bound on $\widetilde{\Lambda} =720$, we obtain a radius
of 13.04 km. Similarly the lower bound of $\widetilde{\Lambda} = 70$ 
gives a radius of $8.85$ km.

It is already known that the tidal deformability is another measure for the
radius of a neutron star. We exploit an alternative method to extract the 
radii of component masses from the upper bound on the combined tidal 
deformability, set by GW170817. In Fig. 5, the combined tidal deformabilities
are plotted as a function of radii corresponding to primary and secondary
masses 1.55 and 1.2 M$_{\odot}$ in right and left panels, respectively, for 
different EoSs as shown in Fig. 1. Curves in both panels are fitted with 
$\widetilde{\Lambda} \propto R^{6}$. The error bars are estimated with 
respect to the fit. The upper bound on the radius of 1.2 
M$_{\odot}$ is 13.08 km whereas that of 1.55 M$_{\odot}$ is 12.99 km. These 
values are in very good agreement with the radius obtained from analytical 
relation given by Eq.({\ref{rad}}) using $\widetilde{\Lambda}=720$. 

\begin{figure}[ht!]
\epsscale{0.60}
\plotone{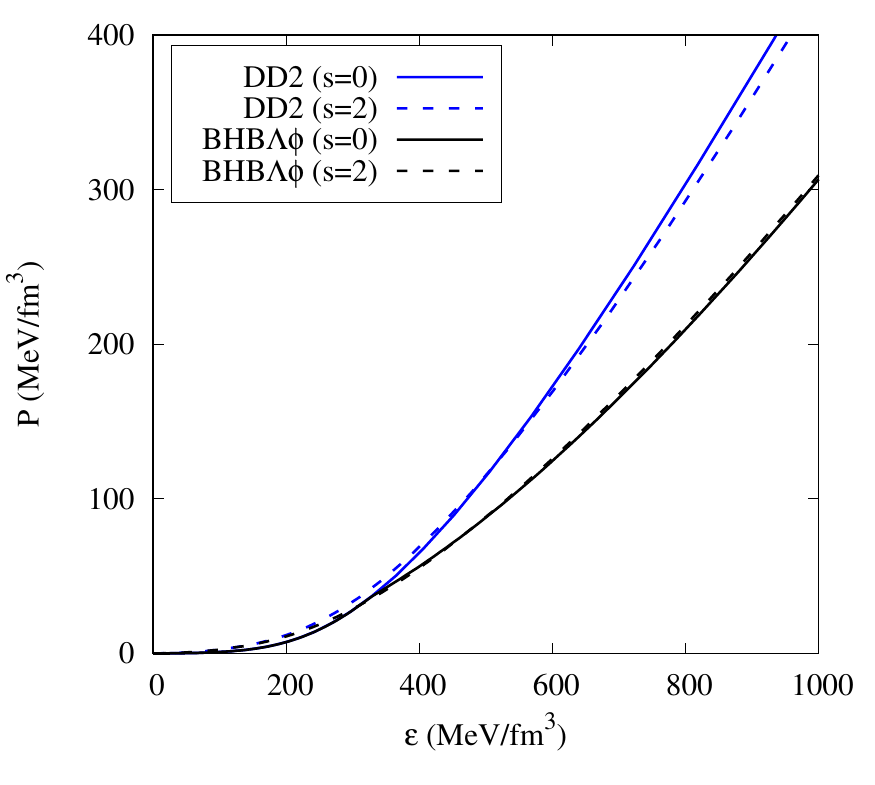}
\caption{Pressure as a function of energy density is plotted for DD2 and 
BHB$\Lambda \phi$ EoS models at entropy per baryon $s=0$ and $s=2$.}
\label{eosfs}
\end{figure}
\begin{figure}
\epsscale{0.60}
\plotone{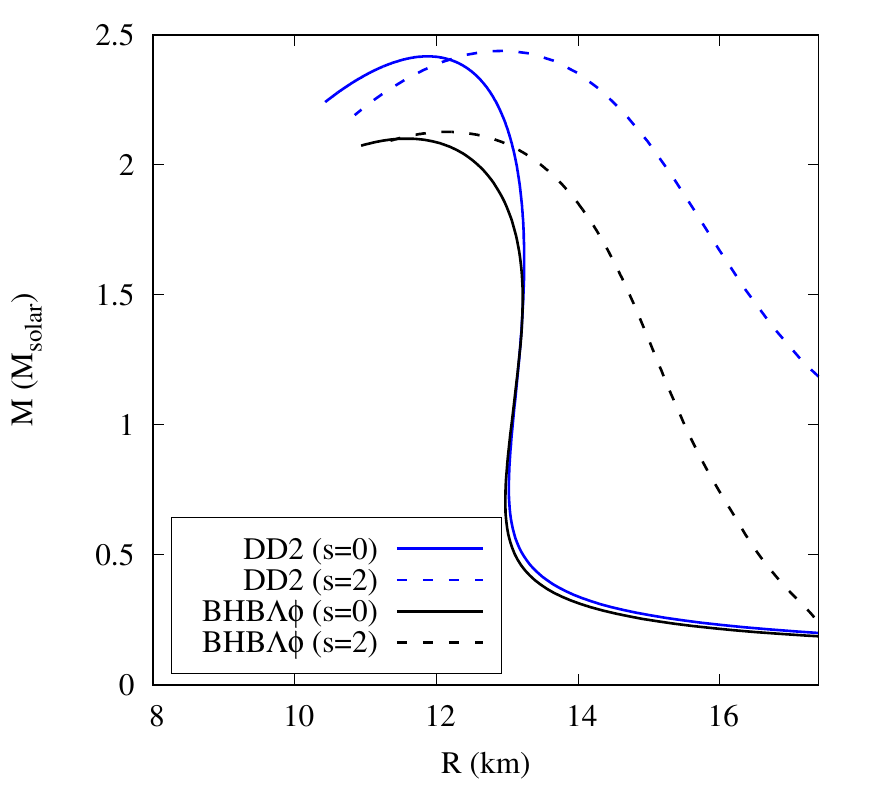}
\caption{Mass-Radius relationships are plotted for DD2 and BHB$\Lambda \phi$
EoSs at entropy per baryon $s=0$ and $s=2$.}
\label{mrfs}
\end{figure}

Next we focus on finite temperature EoSs and its impact on the properties 
of the remnant. We plot the pressure versus energy density and 
mass-radius relationships of non-rotating neutron star sequences for DD2 and
BHB$\Lambda \phi$ EoSs at entropy per baryon $s=0$ and $s=2$ in Figs.
\ref{eosfs} and \ref{mrfs}. We find from Fig.{\ref{mrfs}} that the 
DD2 and BHB$\Lambda \phi$ EoSs for $s=2$ result in higher 
maximum masses of non-rotating neutron stars due to thermal pressure. However, 
the thermal effects on the maximum masses are 
tiny in both cases. Moreover, we observe that radii are much larger for $s=2$ 
than $s=0$ case. The thermal effect on the radius is significantly 
pronounced for DD2 EoS.   

\begin{deluxetable}{rccccccccccccccc}[ht!]
\tablecolumns{17}
\tablewidth{0pc}
\tablecaption{Gravitational mass of rigidly 
rotating remnant at the Kepler frequency, the corresponding baryon mass and 
moment of inertia for entropy/baryon $s=0$ and $s=2$ are given by this table.} 
\tablehead{
EoS& \multicolumn{4}{c}{s=0}  && \multicolumn{4}{c} {s=2} \\
& $M_G^{Rot}$&$M_B^{Rot}$&$I$&$f_{Kep}$&&$M_G^{Rot}$ &$M_B^{Rot}$ &$I$&$f_{Kep}$\\
&$(M_{\odot})$ &$(M_{\odot})$ &($10^{38}$ kg m$^2$)&$(kHz)$&&$(M_{\odot})$&$(M_{\odot})$ &($10^{38}$ kg m$^2$)&$(kHz)$}
\startdata
DD2&2.606&3.004&5.439&1.236&& 2.657&2.998&5.400&1.109\\  
BHB$\Lambda \phi$&2.525&2.914&4.204&1.424&& 2.427&2.717&3.755&1.269\\  
SFHo&2.444&2.856&3.214&1.763&& 2.447&2.807&3.346&1.606\\  
SFHx&2.556&3.000&4.051&1.581&& 2.492&2.832&3.715&1.425\\  
TM1&2.623&3.003&5.241&1.228&& 2.634&3.001&6.767&1.011\\  
TMA&2.439&2.785&4.191&1.315&& 2.448&2.728&4.460&1.099\\  
\hline
\enddata
\label{table1}
\end{deluxetable}
Now we explore the thermal effects on the properties of the remnant. 
Assuming a long lived remnant, the earlier study of maximum rotation rate and
moment of inertia of the remnant was carried out by using large numbers of 
cold and $beta$-equilibrated EoSs \citep{ligo6}. Here we compute the maximum 
rotation rate and moment of inertia of the remnant for different finite 
temperature EoSs using the numerical library LORENE which is best suited for
rapidly rotating compact stars at a fixed entropy per baryon. Though the 
rotation rate of the remnant could exceed the Keplerian limit of the uniformly 
rotating neutron star having the same baryon mass as the remnant because of 
differential rotation, the mass-shedding limit was 
set as the upper limit of the remnant's rotation \citep{ligo6,radice2}. 
It was estimated in Ref.{\citep{ligo6}} that the upper bound 
of the initial baryon mass of the remnant was 3.05 M$_{\odot}$. Some amount of
baryon mass ($< 0.1$ M$_{\odot}$) was ejected from the remnant. In our analysis,
we find that the maximum masses at the Keplerian speed for some EoSs are much 
higher than 3.05 M$_{\odot}$. For those EoSs, we restrict calculations to the
baryon mass of 3 M$_{\odot}$ for the remnant at the mass-shedding limit. We
record the baryon mass, moment of inertia and Keplerian frequencies for EoSs
with fixed entropy per baryon $s=0$ and $s=2$ in Table 2. We find that the 
Keplerian frequencies at $s=2$ for all EoSs are appreciably lower than those
of $s=0$ cases. We note from our discussion on the thermal effects on radii of
neutron stars, as evident in Fig.{\ref{mrfs}}, that neutron stars at higher 
entropy per baryon are bigger in size than that of cold neutron stars. 
This compensates the increase 
in mass due to thermal effects keeping the total baryon mass close to that
of the cold remnant \citep{kaplan}. 
The hot remnant attains the mass-shedding limit at significantly 
lower frequencies than that of the cold remnant. 
Furthermore, this shows that the 
Keplerian frequencies of the remnant using zero temperature EoSs are grossly
underestimated.          

It is worth mentioning here that the thermal effects on the remnant in BNS
merger were earlier explored using EoSs at fixed temperatures \citep{kaplan}. 
This was carried out using a numerical method known as Cook, Shapiro and 
Teukolsky (CST) solver whereas our calculations are performed
in the numerical library LORENE using fixed entropy EoSs. The qualitative 
outcome of an extended remnant and lower Keplerian frequency
due to thermal effects is the same in both cases. Furthermore, we could impose 
the constrain on the total baryon mass of the remnant as obtained
from the analysis of GW170817 in our calculation.

\section{Summary and Conclusions}       
We have investigated the properties of the binary components and remnant of 
GW170817 using zero and finite temperature EoSs constructed within the 
framework of relativistic mean field models. 
The structures and tidal deformabilities of neutron stars involved in the BNS 
merger are estimated using zero temperature EoSs. All EoSs used in the
mass-radius
relationship are compatible with the two solar mass neutron star. On the other
hand, only soft EoSs and mildly stiff EoSs such as SFHo, SFHx, G230a, G240a,
BHB$\Lambda \phi$ are allowed
by 50 $\%$ and 90 $\%$ confidence intervals on the combined tidal 
deformability 
parameter as obtained from the gravitational wave signal of GW170817. It 
is noted that the tidal deformability in GW170817 and the 2 M$_{\odot}$ 
pulsars jointly put strong constraint on the EoS of dense matter. An 
analytic relation is obtained to estimate the radius of neutron stars involved
in the merger using the knowledge of combined tidal deformability. The radius
is $\sim 13$ and 8.85 km for the upper and lower bounds on the combined tidal 
deformability, respectively.       

Next the structures, moment of inertia and Kepler frequency of the rigidly
rotating remnant of GW170817 have been investigated using EoSs at fixed entropy 
value per baryon $s=2$ and compared with those calculated at zero temperature 
i.e. $s=0$. We have taken the baryon mass of the remnant $\lesssim 3 M_{\odot}$
as estimated in Ref.{\citep{ligo6}}. Furthermore, the remnant is rigidly 
rotating at the mass-shedding limit after differential rotation has been
eased out due to the effects of viscosity and magnetic winding. It is 
observed that the thermal effects have
very negligible impact on the mass of the remnant. However, the radius of the 
remnant increases significantly due to the thermal effects compared with the 
situation 
at zero temperature. Consequently, we find that the Kepler frequency is 
appreciably lower for $s=2$ EoSs than those with $s=0$. It may be concluded 
that Kepler frequency calculated with zero temperature is grossly 
overestimated in Ref.{\citep{ligo6}}.     

\acknowledgments

SS acknowledges the support and local hospitality of SINP where a part of this
was completed. DB acknowledges the fruitful discussion with Horst St\"ocker and
hospitality at FIAS during this work.

\software{LORENE, URL package \citep{eric}, CompOSE \citep{composemanual}},
GNU Scientific Library (GSL;\citep{gough})

%%%%%%%%%%%%%%%%%%%%%%%%%%%%%%%%%%%%%%%%%%%%%%%%%%%%%%%%%%%%%%%%%%%%%%%%%%%

\begin{thebibliography}{}
\bibitem[Abbott et al.\ (2017a)] {ligo1}
Abbott, B. P., Abbott, R., Abbott, T. D. et al. 2017, 
PhRvL, 119, 161101
\bibitem[Abbott et al.\ (2017b)] {ligo2}
Abbott, B. P., Abbott, R., Abbott, T. D. et al. 2017, ApJL, 848 L13
\bibitem[Abbott et al.\ (2018)] {ligo3}
Abbott, B. P., Abbott, R., Abbott, T. D. et al. 2018, PhRvL, 121, 161101 
\bibitem[Abbott et al.\ (2019a)] {ligo4}
Abbott, B. P., Abbott, R., Abbott, T. D. et al. 2019, PhRvX, 9, 011001
\bibitem[Abbott et al.\ (2019b)]{ligo5}
The LIGO Scientific Collaboration, The VIRGO collaboration, Abbott, B. P.,
Abbott, R., Abbott, T. D. et al.  2019, ApJ, 875, 160 
\bibitem[Abbott et al.\ (2019c)]{ligo6}
The LIGO Scientific Collaboration, The VIRGO collaboration, Abbott, B. P., 
Abbott, R., Abbott, T. D. et al. 2020, CQGra, 37, 045006 
\bibitem[Antoniadis et al.\ (2013)]{anto} Antoniadis, J., Freire, P. C. C., 
Wex, N. et al. 2013, Science, 340, 448
\bibitem[Banik et al. (2014)]{bhb} Banik, S., Hempel, M., \& Bandyopadhyay, D. 
2014, Astrophys. J. Suppl., 214, 22
\bibitem[Bonazzola et al.\ (1993)]{BGSM}
Bonazzola, S., Gourgoulhon, E., Salgado, M., \& Marck, J.-A. 1993,
Astron. Astrophys. 1993, 278, 421 
\bibitem[Baym et al.\ (1971)]{bps} Baym, G., Pethick, C. J., \& 
Sutherland P. 1971, Astrophys. J.,  170, 299 
\bibitem[Bhat \& Bandyopadhyay (2019)]{sajad} Bhat, S. A., \& Bandyopadhyay, D.
2019, J. Phys. G, 46, 014003 
\bibitem[Breschi et al.\ (2019)]{radice3} Breschi, M., Bernuzzi, S.,
Zappa, F. et al. 2019, PhRvD, 100, 104029 
\bibitem[Char \& Banik (2014)]{char} Char, P., \& Banik, S. 
2014, Phys. Rev. C, 90, 015801
\bibitem[Cioffi et al.\ (2019)]{giacomazzo} Cioffi, R., Kastaun, W., 
Kalinani, J. V., \& Giacomazzo, B. 2019, 
Phys. Rev. D, 100, 023005 
\bibitem[Cromartie et al.\ (2019)]{croma}
Cromartie, H. T., Fonseca, E., Ransom S. M. et al. 2020, NatAs, 4, 72 
\bibitem[De et al.\ (2018)]{de}
De, S., Finstad, D., Lattimer, J. M. et al. 2018, PhRvL, 121, 091102 
\bibitem[Demorest et al.\ (2010)]{demo} 
Demorest P. B., Pennucci, T., Ransom, S. M. et al. 2010, Natur, 467, 1081
\bibitem[Dover \& Gal (1985)]{dov} Dover, C. B.,
\& Gal, A. 1985, Prog. Part. Nucl. Phys., 12, 171 
\bibitem[Endrizzi et al.\ (2019)]{radice1} Endrizzi, A., Perego, A.,
Fabbri, F. M. et al. 2020, EPJA, 56, 15 
\bibitem[Fattoyev et al. \ (2018)]{fattoyev}
Fattoyev, F. J., Piekarewicz, J., \& Horowitz, C. J. 2018,
Phys. Rev. Lett., 120, 172702 
\bibitem[Fischer et al.\ (2014)]{fis} Fischer, T., Hempel, M., Sagert, I., 
Suwa, Y., \& Schaffner-Bielich, J.2014, Eur. Phys. J. A, 50, 46 
\bibitem[Fujibayashi et al.\ (2018)]{fuji} 
Fujibayashi, S., Kiuchi, K., Nishimura, N., Sekiguchi, Y., \& Shibata, M. 2018, 
Astrophys. J., 860, 64
\bibitem[Gough et al. \ (2009)]{gough}
Gough, B. 2009, GNU Scientific Library Reference Manual (3rd ed; Network Theory
Ltd.)
\bibitem[Gourgoulhon et al. \ (2016)]{eric}
Gourgoulhon, E., Grandcl\'ement, P., Marck, J.-A., Novak, J., \& Taniguchi, K.
2016, LORENE spectral methods differential equation solver,
Astrophysics Source Code Library, ascl:1608.018
\bibitem[Hanauske et al.\ (2019)]{magic} 
Hanauske M., Bovard, L., Steinheimer, J. et al. 2019, J. Phys. Conf. Ser., 
1271, 012023
\bibitem[Hebeler et al.\ (2013)]{heb} Hebeler, K., Lattimer, J. M., 
Pethick, C. J., \& Schwenk, A. 2013, 
Astrophys. J., 773, 11 
\bibitem[Hempel et al.\ (2012)]{hem12} Hempel, M., Fischer, T., 
Schaffner-Bielich, J., \& Liebend\"orfer, M. 2012, 
Astrophys. J., 748, 70 
\bibitem[Hempel \& Schaffner-Bielich (2010)]{hs1} Hempel, M.,
\& Schaffner-Bielich, J. 2010, Nucl. Phys. A, 837, 210
\bibitem [Hewish et al.\ (1968)]{bell}
Hewish, A., Bell, S. J., Pilkington J. D. H. et al. 1968,
Nature, 217, 709 
\bibitem[Hornick et al.\ (2018)]{horn} Hornick, N., Tolos, L., Zacchi, A.,
Christian, J.-E., \& Schaffner-Bielich, J. 2018, Phys. Rev. C, 98, 065804 
\bibitem[Hotokezaka et al.\ (2013)]{kenta} 
Hotokezaka, K., Kiuchi, K., Kyutoku, K., Muranushi, T., Sekiguchi, Y., Shibata, M., \& Taniguchi, K. 2013, Phys. Rev. D, 88, 044026 
\bibitem[Kiuchi et al.\ (2018)]{kiuchi} 
Kiuchi, K., Kyutoku, K., Sekiguchi, Y., \& Shibata, M. 2018, 
Phys. Rev. D, 97, 124039 
\bibitem[Lattimer \& Lim (2013)]{jim} Lattimer, J. M., \& Lim, Y. 2013,
Astrophys. J., 771, 51
\bibitem[Lonardoni et al.\ (2019)]{lona} 
Lonardoni, D., Tews, I., Gandolfi, S., \& Carlson, J. 2019, 
arXiv:1912.09411
\bibitem[Margalit \& Metzger \ (2017)]{metz}
Margalit, B., \&  Metzger, B. D. 2017,
Astrophys. J. Lett., 850, L19 
\bibitem[Mares et al.\ (1995)]{mar} Mares, J., 
Friedman, E., Gal, A., \& Jennings, B. 1995, Nucl. Phys. A, 594, 311
\bibitem[Marques et al.\ (2017)]{marq} Marques, M., Oertel, M., Hempel, M. \& 
Novak, J. 2017, Phys. Rev. C, 96, 045806
\bibitem[Mellinger et al.\ (2017)]{weber}
Mellinger Jr, R. D., Weber, F., Spinella, W., Contrera G. A., \& 
Orsaria, M. G. 2017, Universe, 3, 1 
\bibitem[Millener et al.\ (1988)]{mil} Millener, D. J., Dover, C. B.,
\& Gal, A. 1988, Phys. Rev. C, 38, 2700
\bibitem[Miller et al.\ (2019)]{miller} 
Miller, M. C., Lamb, F. K., Dittmann, A. J. et al. 2019, ApJL, 887, L24 
\bibitem[Most et al.\ (2019)]{most} Most, E. R., Papenfort, L. J.,
Dexheimer, V et al. 2019, PhRvL, 122, 061101 
\bibitem[Most et al.\ (2018)]{sch18}
Most, E. R., Weih, L. R., Rezzolla, L., \& Schaffner-Bielich, J. 2018,
Phys. Rev.  Lett., 120, 261103 
\bibitem[Negele \& Vautherin (1973)]{neg} Negele, J. W., \& Vautherin, D.
1973, Nucl. Phys.  A, 207, 298 
\bibitem[Oertel et al.\ (2017)]{rmp} Oertel, M., Hempel, M., Kl\"ahn, T., \&
Typel, S. 2017, Rev. Mod. Phys., 89, 015007 
\bibitem[Perego et al.\ (2019)]{perego}
Perego, A., , Bernuzzi, S., \& Radice, D. 2019, 
Eur. Phys. J. A, 55, 124 
\bibitem[Radice et al.\ (2018a)]{radice2} Radice, D., Perego, A., Bernuzzi, S.,
\&  Zhang, B. 2018, 
Mon. Not. R. Astron. Soc, 481, 3670 
\bibitem[Radice et al.\ (2018b)]{radice4}
Radice, D., Perego, A., Zappa, F., \& Bernuzzi, S. 2018, 
Astrophys. J. Lett., 852, L29 
\bibitem[Radice et al.\ (2017)]{ott} 
Radice, D., Bernuzzi, S., Del Pozzo, W., Roberts L. F., \& Ott, C. D. 2017,
Astrophys. J., 842, L10 
\bibitem[Raithel et al.\ (2018)]{ozel}
Raithel, C., \"Ozel, F., \& Psaltis, D. 2018,
Astro. Phys. J., 857, L23 
\bibitem[Rezzolla et al.\ (2018)]{luci}
Rezzolla, L., Most, E. R., \& Weih, L. R. 2018,
Astrophys. J. Lett., 852, L25
\bibitem[Riley et al.\ (2018)]{riley} 
Riley, T. E., Watts, A. L., Bogdanov, S. et al. 2019, ApJL, 887, L21 
\bibitem[Ruiz et al.\ (2018)]{shapiro}
Ruiz, M., Shapiro, S. L., \& Tsokaros, A. 2018, 
Phys. Rev. D, 97, 021501 
\bibitem[Schaffner \& Mishustin (1996)]{sch} 
Schaffner, J., \& Mishustin, I. N. 1996, Phys. Rev. C, 53, 1416
\bibitem[Schaffner et al.\ (1992)]{sch92}
Schaffner, J., St\"ocker, H., \& Greiner, C. 1992, Phys. Rev. C, 46, 322
\bibitem[Sekiguchi et al.\ (2011)]{seki} 
Sekiguchi, Y., Kiuchi, K., Kyutoku, K., \& Shibata, M. 2011, 
Phys. Rev.  Lett., 107, 051102 
\bibitem[Shibata et al.\ (2017)]{shibata}
Shibata, M., Fujibayashi, S., Hotokezaka, K. et al. 2017, PhRvD, 96, 123012 
\bibitem[Shibata et al.\ (2019)]{enping}
Shibata, M., Zhou, E., Kiuchi, K., \& Fujibayashi, S. 2019, 
Phys. Rev. D, 100, 023015 
\bibitem[Steiner et al.\ (2005)]{stein05} Steiner, A. W., Prakash, M., 
Lattimer, J. M., \& Ellis, P. J. 2005, 
Phys. Rep., 411, 325 
\bibitem[Steiner et al.\ (2013)]{stein13} Steiner, A. W., Hempel, M., \&
Fischer, T. 2013, Astrophys. J., 774, 17 
\bibitem[Steiner et al.\ (2010)]{stein10} Steiner, A. W., Lattimer, J. M., \&
Brown, E. F., Astrophys. J., 722, 33 
\bibitem[Stone et al.\ (2014)]{stone} Stone, J. R., Stone, N. J., \&
Moszkowski, S. A. 2014, Phys. Rev. C, 89, 044316
\bibitem[Sugahara \& Toki (1994)]{toki} 
Sugahara, Y., \& Toki, H. 1994, Nucl. Phys. A, 579, 557
\bibitem[Tews et al.\ (2017)]{tews1} 
Tews, I., Lattimer, J. M., Ohnishi, A., \& Kolometsev, E. E.  2017, 
Astrophys. J., 848, 105
\bibitem[Toki et al.\ (1995)]{toki95} 
Toki, H., Hiratai, D., Sugahara, Y., Sumiyoshi, K., \& Tanihata, I.  1995, 
Nucl. Phys. A, 588, c357
\bibitem[Typel \& Wolter (1999)]{wol} Typel, S., \& Wolter, H. H. 1999, Nucl.
Phys. A, 656, 331
\bibitem[Typel et al.\ (2010)]{typ10} Typel, S., R\"opke, G., Kl\"ahn,
T., Blaschke, D., \& Wolter, H. H. 2010, Phys. Rev. C, 81, 015803
\bibitem[Typel et al.\ (2013)]{composemanual} Typel, S., Oertel, M., \& 
Klaehn, T. 2013, arXiv:1307.5715
\bibitem[Zhao \& Lattimer (2018)]{zhao}
Zhao, T. \& Lattimer, J. M. 2018, Phys. Rev. D, 98, 063020 
\bibitem[Kaplan et al.\ (2014)]{kaplan} Kaplan, J. D., Ott C. D., 
O'Connor, E. P., Kiuchi, K., Roberts, L., \& Duez M. 2014, Astrophys. J., 
790, 19 

\end{thebibliography}
\end{document}